# Easy Java/JavaScript Simulations as a tool for Learning Analytics

**Francisco ESQUEMBRE**[a*], **Félix J. GARCÍA CLEMENTE**[b], **Rafael CHICÓN**[c], **Lawrence WEE**[d], **Leong Tze KWANG**[d] & **Darren TAN**[d]
[a]Department of Mathematics, University of Murcia, Spain
[b]Department of Computer Engineering and Technology, University of Murcia, Spain
[c]Department of Electromagnetism and Electronics, University of Murcia, Spain
[d]Ministry of Education, Singapore
*fem@um.es

**Abstract:** In this paper we introduce the new and planned features of Easy Java/JavaScript Simulations (EJS) to support Learning Analytics (LA) and Educational Data Mining (EDM) research and practice in the use of simulations for the teaching and self-learning of natural sciences and engineering. Simulations created with EJS can now be easily embedded in a popular Learning Management System using a new plug-in that allows creation of full-fledged instructional units that also collect and record fine-grained, instructional-savvy data of the student's interaction with the simulation. The resulting data can then be mined to obtain information about students' performance, behaviors, or learning procedures with the intention to support student learning, provide instructors with timely information about student performance, and also help optimize the pedagogic design of the simulations themselves. We describe the current development and architecture, as well as future directions for testing and extending the current capabilities of EJS as a modelling and authoring tool to support LA and EDM research and practice in the use of simulations for teaching science, in particular in the context of the increasingly popular on-line learning platforms.

## 1. The need for Learning Analytics for simulations

Teaching science and engineering with simulations (also known as virtual laboratories) is common ground nowadays (Christian et al. 2015) (Perkins et al. 2006). The combined benefits of the visualization and interaction capabilities of a well-designed simulation with the use of a correct pedagogical approach has been the subject of much educational research. The literature offers plenty of references for this successful topic and, consequently, there exist also several collections of ready-to-use simulations and associated curricular material, mainly in the fields of Physics and Control Engineering (OSP 2019) (PhET 2019) (Sáenz et al. 2015).

However, most of this educational research was conducted in controlled classrooms, where typically a not too large number of students used the materials under the direct supervision of the teacher. The proliferation of online courses, of which the most prominent examples are Massive Online Open Courses (MOOC), has opened new ways for instructors to use course materials, including simulations, with a large number of students and in scenarios where students may not be under the direct supervision of a teacher, leaving room for self-exploration and a growing need to consider student's own learning strategies and procedures (Dervić et al. 2018) (Chamberlain et al. 2014). It is not clear, at this moment, that simulations designed for classroom use can be used with the same success in on-line, perhaps unsupervised, courses. This new ground constitutes both a challenge and a whole new world of research possibilities. And it comes with an entry bonus: tons of data.

Since the very beginning, MOOC and other on-line pedagogic platforms, such as Learning Management Systems (LMS), have been a great source for those seeking feedback from the students' use of the curricular material at their disposal (Breslow et al 2013). Software clients (programs running on students' computers or devices acting as interfaces to the material) can capture information about when and how students access the data. Computer servers providing the material can then store this data and offer them back to researchers, together with other relevant personal or contextual information about students themselves.

This new and large availability of data has fueled research fields that measure and analyze this data with the purpose of monitoring, supporting, understanding and optimizing student learning, learning materials and even the learning environments. Two of these fields are Learning Analytics (LA) and Educational Data Mining (EDM), which are actually closely related. The differences between them are whether researchers have a more human perspective in their approaches and methods (LA), or rely more on automatic discovery (EDM) (Viberg et at. 2018). (We may, however, refer in this paper to both approaches just as *learning analytics*.)

However, although collecting general information from user interaction with the materials on an LMS or MOOC is relatively easy, it is not always clear that an *all-the-information-we-can-get* policy is the most suitable form to collect the most relevant *features* for applying machine learning algorithms or statistical studies that will produce correct predictions on the retention and performance of students or, better yet, recognize improved students' learning outcomes. More and more authors now claim that empirical LA and EDM studies should always be carried out in the context of a sound theoretical framework of how learners learn. Only through the standard scientific process of hypothesis-validation within a theory can concrete exploration findings be converted into accepted explanations and lead to sensible pedagogic actions of general applicability (Wise & Shaffer 2015) (Gašević et al 2015) (Fincham et al. 2019).

An analysis of the different methods applied to data collected from online sessions revealed that predictive methods (those trying to predict students' retention and grades) have been dominating the field, at least until 2016. But researchers are currently shifting towards methods seeking a deeper understanding of student's learning experiences (Viberg et at. 2018). As a consequence, more research in being directed towards considering learning as a process, one in which learning *events* form a sequence of cognitive operations or *patterns* leading to possible learning outcomes (Winne et al. 2019), with some authors claiming that "while learning trace data provide granular details about students' realized intentions, there is an uncertainty in how to connect patterns in traces of digital behavior with the features of the learning process" (Jovanović et al. 2019). Clearly, much research lies ahead.

This is where we think we can help. Easy Java/JavaScript Simulations (EJS) is a free software modeling and authoring tool designed to help instructors design, implement and deploy computer simulations of scientific and engineering processes. EJS was conceived and developed with the goal of putting the power of creating simulations directly in the hands of teachers and educational researchers, a process that we refined during years of close work with them. EJS facilitates the numerical, computing, graphical user interface building, and deployment tasks of developing simulations so that instructors and curricula developers can concentrate on the pedagogical aspects: how to model a given phenomenon, how to visualize it, what interaction to offer to students, how to use the simulation in a given pedagogical context…

The result is an award-winning software tool that uses a standardized structure, both simple and powerful, for the creation of a simulation, in particular for programming its model and designing its user interface or view (Esquembre 2004). This standardization has permitted the creation of hundreds of educational simulations, mostly in the field of Physics and in Control Engineering (Christian et al. 2011) (Esquembre 2015) (Wee et al. 2015), that can not only be used 'out-of the-shelf' in educational webs or LMS, but that are also adoptable and adaptable. That is, their source code can be easily inspected by instructors other than the author and even modified to custom-tailor them to particular needs.

Our approach to turning EJS into a tool for helping conduct Learning Analytics and Educational Data Mining research and practice is based both on the standardized structure of the simulations created with EJS and in the vast number of examples already available, particularly for Physics instructions in the Open Source Physics (OSP) collection of the ComPADRE library (OSP 2019) and the Singapore collection of OSP simulations (OSP@SG 2019). Downloading and inspecting the simulation with EJS allows the instructor to become familiar with the pedagogic features of the simulation. In particular, with the different user interface controls (*view elements,* in EJS dialect). Recompiling the simulation with the latest version of EJS automatically produces a standard HTML simulation that can be readily deployed via any web page. Moreover, using a dedicated plug-in of our creation, the simulation can be readily included in any Moodle course that, when operated by the user, will collect data of the user interaction with the simulation. Moodle is a very popular LMS (Moodle 2019).

Besides collecting *classical* data of the student interaction with the LMS or the simulation (such as student ID, activity, start time, end time, idle time, mouse clicks and drags, keystrokes…) which can be used for statistical or even machine learning analysis and prediction (Hussain et al. 2019), the EJS generated simulation also collects information of the view element that was interacted with. Since the instructor knows the action invoked by this view element, she or he can use this information for more fine-grained study of the learning event that was intended by the student. (For instance, to recognize that the student is running the simulation with different sets of initial conditions, or attempting to visualize the system in different forms.) Additionally, the instructor can modify the simulation to add to each interaction record information about particular model variables, or slightly change the model of the simulation to automatically generate custom records at particular times or states. (For example, to distinguish idle time from time when the student is observing the simulation while it runs.) The instructor will later collect and study this custom information according to her or his pedagogic interest, (For instance, searching for inquiry-based learner's strategies).

We believe that the combination of these features can help EJS users to create a number of different scenarios for conducting LA and EDM research about the use of simulations in the teaching and learning of science and engineering in on-line, perhaps unsupervised learning platforms. We plan to conduct such a study ourselves with a number of teachers and students of Physics, both as a proof of concept of the tool, but also to fine-tune the amount and diversity of information that EJS generated simulations should offer to researchers, and to study how a particular learning analytic approach can help assist and improve student learning in this context, and also help improve the simulation design itself.

The rest of the paper is organized as follows. Section 2 explains briefly the architecture of EJS to show where LA and EDM researchers can look and edit an existing simulation to learn and decide about the user interaction information collected. Section 3 describes the learning analytics scenario now possible with EJS simulations. Section 4 shows the Moodle plug-in we have created to install the simulation in a Moodle course and collect the data generated by it, together with a simple dashboard that uses the collected information for teacher supervision of students' participation. Finally, Section 5 draws some conclusions and describes future work.

## 2. EJS architecture

EJS follows a Model-View-Controller software design pattern and provides developers with a corresponding simplified interface for creating a simulation. Using this interface, authors (typically science or engineering instructors) specify *at a high level* the model of a simulation (i.e. the set of variables that defines the system under study, and the equations and algorithms that specify how it evolves in time o reacts to user interaction), and its view (the graphical user interface that allows students to visualize the state of the system and to interact with it to control the execution of the simulation or to change the value of variables).

Specifying the model requires a little bit of programming, although EJS facilitates standard tasks such as organizing the computational flow of the simulation, declaration of variables and numerical processes such as solving ordinary differential equations (ODE). (See Figure 1.)

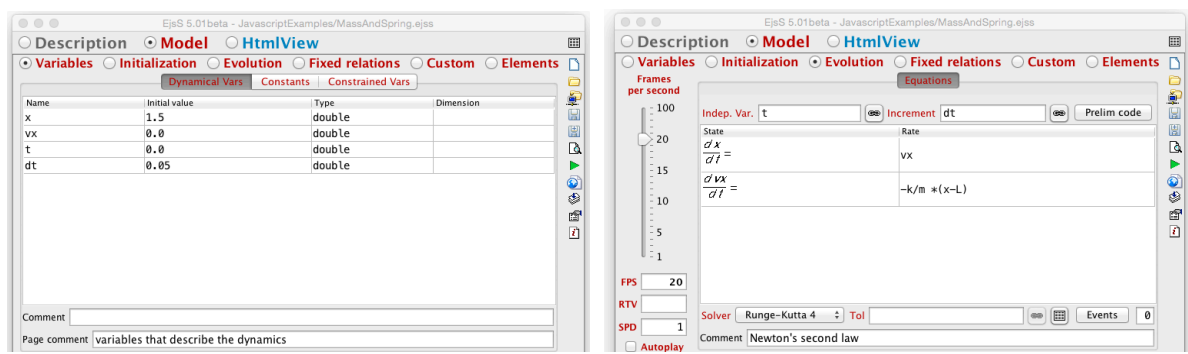

*Figure 1.* Declaration of variables (left) and use of the ODE editor (right) for the model of a mass and spring system.

To design the view of the simulation, instructors select from a wide palette of *view elements*, individual graphic components each specialized in a given type of visualization or user interaction, to build a tree of elements that turns into a sophisticated HTML interface. A typical view contains a button to reset the simulation to its initial state, fields and sliders to set those initial conditions or modify parameters, and buttons to play, pause, or run the simulation step-by-step. The view also includes animated graphic elements that move in a two- or three-dimensional canvas, providing a virtual representation of the phenomenon being simulated or graphs of data generated as the simulation evolves in time (Figure 2). Model variables are linked to view elements properties to make both the view visualize the state of the system and its evolution, and the model respond to user interaction.

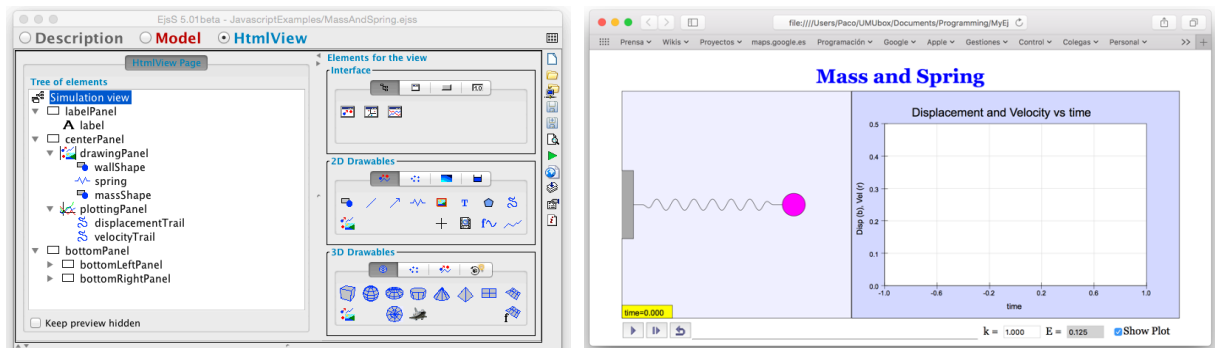

*Figure 2.* Tree and palette of view elements (left) and final HTML user interface (right) for the mass and spring system.

Creating a simulation from scratch requires some training (a three-days workshop can be enough to get a good start). However, inspecting a simulation created by another instructor is straightforward. And modifying the simulation slightly to include new view elements for visualization or control is also relatively simple. Once the simulation is designed, a simple click generates automatically all the code required for a full-fledged simulation that uses a supportive advanced JavaScript library (Garcia Clemente et al. 2014). The generated simulation can then be run in any web browser with standard HTML support on any kind of computer or device, including smartphones or tablets.

## 3. Learning Analytics scenario with EJS simulations

When an EJS generated simulation runs, the underlying library manages all the design information provided by the author within the standard architecture. In particular, the library handles every student interaction with the view in order to execute the corresponding actions to the model (resetting the simulation, stepping it in time, changing a given model variable or parameter…) and can now be instructed to send this interaction information to an external repository in real-time. By default, the system identifies three types of interactions:

- *Mouse and touch events*. These events allow to know how users interact with the simulation. The event information is composed of position, timestamp and target.
- *Actions on view elements*. Every view element has one or more predefined action properties that can be triggered by user interaction. Perhaps the most common ones are the *OnClick* action triggered when pushing a button and the *OnDrag* action triggered when dragging an interactive 2D element. When an action is fired, all the information related to the action is registered.
- *Actions on the model*. The allowed actions on the model are play, pause, step, reset and initialize.

The collected information with these interactions alone is so detailed that it is possible to reproduce the user session, keeping the original timing, like in a video player. This can be of interest when studying the behavior of a given user. However, this information alone cannot be sufficient for some kind of automated studies and may be overkill in others. For example, for a LA study, it could be interesting to know the variables of certain variables (f. i. the initial conditions and parameters) every time the student plays the simulation. Or the value of the time when she or he stops it. This could lead to guesses about a possible student's learning strategy. On the other hand, a machine learning algorithm used in EDM could get too many features using all the information contained in very click, leading to overfitting and therefore poor generalization.

For this reason, EJS JavaScript object-oriented library provides a built-in object called _recorder_ that can be used to configure the behavior of the information recorded for each interaction. This object has a number of functions that can be called from within the model of the simulation to fine-tune what information is sent, and when it is sent, to the external repository. For instance, in the mass and spring example displayed in Figures 1 and 2, the *_recorder* could be instructed to send the value of the model variables *x*, *vx* and *t* (but not all others), every time an interaction happens. Or it could be instructed to send a non-interaction triggered record whenever a given programmatic condition takes places (for instance whenever the graph reaches a maximum). We can only now guess the purpose and needs of the LA or EDM researcher as she or he studies the use of a given simulation by a large number of students. This is perhaps the main advantage of the use of EJS simulations in learning analytics: researchers can customize the information to obtain from user interaction by inspecting the simulation (perhaps one of the many hundred already existing in public libraries) and fine-tuning the captured information using the *_recorder* object.

EJS simulations need other agents in order to collect, store and manage the information thus generated to provide learning analytics facilities to researchers. In the first place, a server needs to serve the simulation HTML page to the student, preferably as part of a curricular unit or course. In the second place, a Learning Record Store (LRS) is needed to store all user interactions produced by the simulation in real-time using a specific repository. Learning Management Systems can help satisfy these two requirements. Instructors can use an LMS to create a training course or program that includes the simulation, together with other accompanying material (theory reviews, instructions, tests…). And an LRS can be integrated into an LMS, capturing the interaction information together with other, particular information about the user (student level, use of other resources, test grades…). See Figure 3.

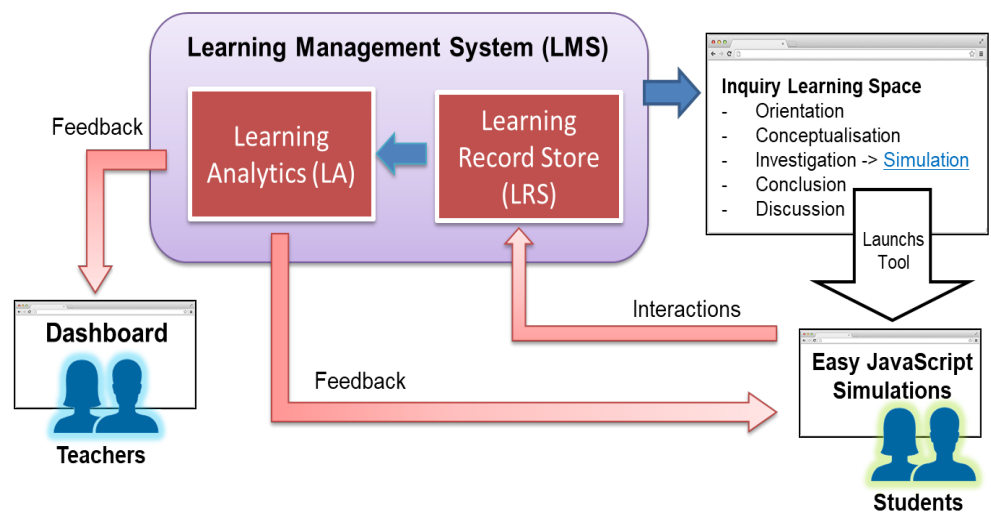

*Figure 3*. Learning Analytics scenario with EJS simulations.

The use of the feedback thus obtained is now a question of choice by the LA or EDM researcher. EJS has always been *pedagogically neutral* (that is, it neither presumes nor fosters any particular pedagogic approach) and will remain so also for the analytic support now offered. This open-minded approach is also responsible for the wide adoption of EJS by instructors, and our intention is to collaborate with researchers interested to expand the capabilities of the *_recorder* object as new features are requested. We expect that statistical and machine learning analysis and prediction methods will be applied to this information and there will be a need to refine features extraction from the interaction of students with the simulation.

## 4. Learning Analytics Moodle extension for EJS simulations

We have created a first development to evaluate and validate the previous scenario using Moodle as LMS (Garcia Clemente et al. 2019). The Moodle extension (plugin) that we have developed monitors and records, in real-time, all the standard student interactions with the EJS simulation. This information is displayed on the instructor's dashboard with a metric of each individual student degree of being "on task" and also shows students' current state on the simulation (Figure 4).

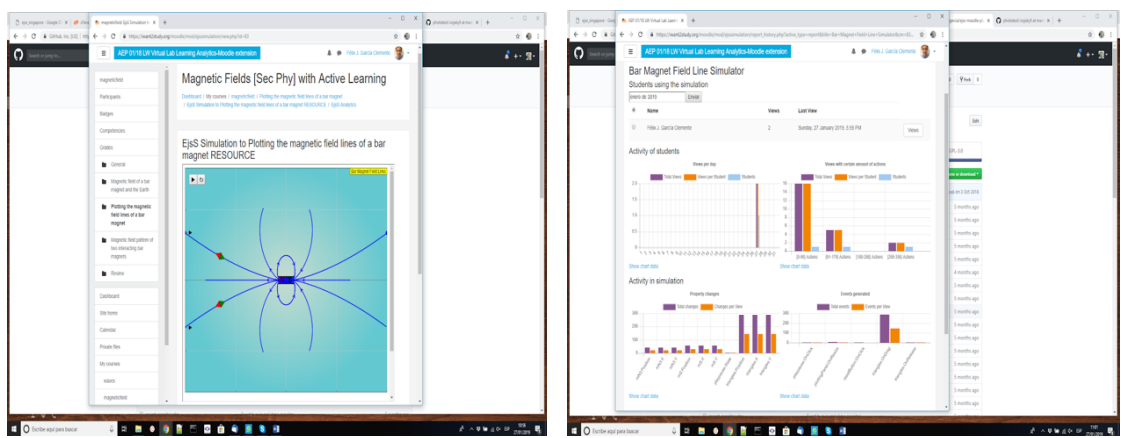

*Figure 4.* An EJS Simulation (left) and the Student Interaction Summary displayed in Moodle (right).

The Moodle plugin[1] is installed on the Open Source Physics at Singapore Moodle server[2], and provides learning analytics capability for teachers running courses based on EJS simulations. This data-rich virtual laboratory functionality can be extended to existing EJS simulations by regenerating the simulation from the source code with version 6.0 of the EJS authoring toolkit, installing our Moodle plugin, and using the simulation in a Moodle course.

## 5. Conclusion and future work

We believe that there is a need for conducting research on the design and use of simulations in the context of online, perhaps unsupervised student learning. This research can be based on techniques of Learning Analytics and Educational Data Mining, always in close connection with the theoretical findings on the use of simulations for teaching natural sciences and engineering.

We wanted to offer support for conducting this research and have adapted the EJS authoring tool to generate simulations that can collect data from student interactions in a customizable, fine-grained, instructional-savvy form, so that researchers can choose what features they want to collect for human-directed or automated studies. We have also created a Moodle plugin to integrate the data collection of simulations with a Learning Record Store and have also provided a first, basic teacher dashboard for monitoring students' activity.

Besides working on improving current facilities in cooperation with interested researchers in a neutral way (without imposing any particular type of analysis or pedagogic approach), we plan to conduct our own research, applying machine learning techniques to detect students not just at risk of dropping or failing a course, but also those not producing correct learning outcomes, while at the same time distilling information on how to improve the simulations design to care for them.

## Acknowledgements

Francisco Esquembre was funded by the Spanish Ministry of Economy and Competitiveness under Project ENE2015-64914-C3-2-R, and the Spanish Ministry of Science, Innovation and Universities under Project MTM2017-84079-P.

## References

Breslow, L., Pritchard, D.E., DeBoer, J., Stump, G.S., Ho, A.D. & Seaton, D.T. (2013). Studying Learning in the Worldwide Classroom: Research into edX's First MOOC. *Research & Practice in Assessment*, 8:13-25.
Chamberlain, J.M., Lancaster, K., Parson, R. & Perkins, K.K. (2014). How guidance affects student engagement with an interactive simulation. *Chemistry Education Research and Practice*, 15:628-638.
Christian, W., Esquembre, F. & Barbato, L. (2011). Open Source Physics. *Science*, 334:1077–1078.

---

[1] The plugin can be obtained freely from https://github.com/felixgarcia/ejss-moodle-plugin

[2] https://iwant2study.org/moodle/

Christian, W., Belloni, M., Esquembre, F., Mason, B. A., Barbato, L. & Riggsbee, M. (2015). The Physlet Approach to Simulation Design. *The Physics Teacher*, 53:419. https://doi.org/10.1119/1.4931011

ComPADRE (2019) Compadre Resources for Services for Physics Education. https://www.compadre.org. Accessed: 2019-08-14.

Dervić, D., Glamočić, D.S., Gazibegović-Busuladžić, A., & Mešić, V. (2018). Teaching physics with simulations: teacher-centered versus student-centered approaches. *Journal of Baltic Science Education*, 17(2), 288-299.

Esquembre, F. (2004). Easy Java Simulations: a software tool to create scientific simulations in java. *Computer Physics Communications*, 156 (2), 199-204.

Esquembre, F. (2015). Facilitating the Creation of Virtual and Remote Laboratories for Science and Engineering Education. *IFAC-PapersOnLine*, 48 (29), 49-58.

Fincham, E., Whitelock-Wainwright, A., Kovanović, V., Joksimović, S., van Staalduinen, J.P., & Gašević, D. (2019). Counting Clicks is Not Enough: Validating a Theorized Model of Engagement in Learning Analytics. In C. Brooks, R.Ferguson & U. Hoppe (Eds), *Proceedings of LAK'19* (pp. 501-510). New York, NY, USA: ACM Press.. https://doi.org/10.1145/3303772.3303775

Garcia Clemente, F.J. & Esquembre, F. (2014). EjsS: A javascript library and authoring tool which makes computational-physics education simpler. *XXVI IUPAP Conference on Computational Physics* (CCP), Boston, USA.

Garcia Clemente, F.J., Loo Kang, W., Esquembre, F., Tze Kwang, L. & Darren, T. (2019). Development of Learning Analytics-Moodle Extension for Easy JavaScript Simulation (EjsS) Virtual Laboratories. *GIREP-MPTL Conference*, Budapest, Hungary.

Gašević, D., Dawson, S. & Siemens, G. (2015). Let's not forget: Learning analytics are about learning. *TechTrends*, *59(1)*. DOI: 10.1007/s11528-014-0822-x

Hussain, M. & Abidi, R. Hussain, M., Zhu, W., Zhang, W., Muhammad, S. & Abidi, R. (2019). Using machine learning to predict student difficulties from learning session data. *Artificial Intelligence Review*, *52 (1)*, 381-407. https://doi.org/10.1007/s10462-018-9620-8

Jovanović, J., Gašević, D., Pardo A., Dawson, S. & Whitelock-Wainwright, A. (2019). Introducing meaning to clicks: Towards traced-measures of self-efficacy and cognitive load. In C. Brooks, R. Ferguson & U. Hoppe (Eds), *Proceedings of LAK'19* (pp. 511-520). New York, NY, USA: ACM Press. https://doi.org/10.1145/3303772.3303782

Moodle (2019) Moodle site. https://moodle.org. Accessed: 2019-08-15.

OSP (2019) Open Source Physics project. http://www.compadre.org/osp. Accessed: 2019-08-14.

OSP@SG (2019) Open Source Physics @ Singapore (2019) https://iwant2study.org/ospsg. Accessed: 2019-08-14.

PhET (2019) PhET Interactive Simulations for Science and Math. https://phet.colorado.edu. Accessed: 2019-08-14.

Perkins, K. Adams, W., Dubson, M., Finkelstein, N., Reid, S., Wieman, C. & LeMaster, R. (2006). PhET: Interactive Simulations for Teaching and Learning Physics. *The Physics Teacher*, *44(1)*, 18-23.

Sáenz, J., Chacón, J., De la Torre, L., Visioli, A. & Dormido, S. (2015). Open and Low-Cost Virtual and Remote Labs on Control Engineering. *IEEE Access*, *3*, 805-814.

Viberg, O., Hatakka, M., Bälter, O. & Mavroudi, A. (2018). The current landscape of learning analytics in higher education. *Computers in Human Behavior*, *89*, 98-110.

Wee, L. K., Lee, T. L., Chew, C., Wong, D., & Tan, S. (2015). Understanding resonance graphs using Easy Java Simulations (EJS) and why we use EJS. *Physics Education*, 50(2), 189-196. doi:10.1088/0031-9120/50/2/189

Winne, P.H., Teng, K., Chang, D., Lin, M.P-C., Marzouk, Z., Nesbit, J.C., Patzak, A., Raković, M, Samadi, D. & Vytasek, J. (2019). nStudy: Software for Learning Analytics about Processes for Self-Regulated Learning. *Journal of Learning Analytics*, *6(2)*, 95-105. https://doi.org/10.18608/jla.2019.62.7

Wise, A. F. & Shaffer, D. W. (2015). Why theory matters more than ever in the age of Big Data. *Journal of Learning Analytics*, *2(2)*, 5-13.